\documentclass[journal,twoside]{IEEEtran}
\usepackage[nocompress]{cite}
\usepackage{color}
\usepackage{amsfonts,amssymb}
\usepackage{amsmath,bm}
\usepackage{ifpdf}
\usepackage{epsfig}
\usepackage{fmtcount}
\usepackage[T1]{fontenc}
\usepackage{balance}
\usepackage{xcolor}
\usepackage{multirow}

\usepackage{array}
\usepackage{eqparbox}
\usepackage{color}
\usepackage{graphicx}
\usepackage{url} 
\usepackage{amsmath}
\usepackage{makecell}
\usepackage{balance}
\usepackage{tabu}
\usepackage{algorithm}
\usepackage{algorithmic}
\usepackage{longtable}

\usepackage{pifont}
\usepackage[utf8]{inputenc}
\usepackage[T1]{fontenc}
\usepackage[printonlyused]{acronym}
\usepackage{subcaption}
\usepackage[flushleft]{threeparttable}

\usepackage{xcolor}

\acrodef{MTC}{machine type communication}
\acrodef{LTE}{Long Term Evolution}
\acrodef{IoT}{internet of things}
\acrodef{NB-IoT}{Narrowband IoT}
\acrodef{eNB}{E-UTRAN Node B}
\acrodef{CID}{cell ID}
\acrodef{DI}{deep indoor}
\acrodef{I}{indoor}
\acrodef{OW}{outdoor walking}
\acrodef{OD}{outdoor driving}
\acrodef{QoS}{quality of service}
\acrodef{RSRP}{reference signal received power}
\acrodef{RSRQ}{reference signal received quality}
\acrodef{RSSI}{received signal strength indicator}
\acrodef{SINR}{signal-to-interference-plus-noise ratio}
\acrodef{PCI}{physical cell identifier}
\acrodef{EARFCN}{E-UTRA absolute radio frequency channel number}
\acrodef{MCC}{mobile country code}
\acrodef{MNC}{mobile network code}
\acrodef{GPS}{global positioning system}
\acrodef{LoS}{line of sight}
\acrodef{BTS}{base transceiver station}
\acrodef{RA}{random access}
\acrodef{CL}{coverage level}
\acrodef{ICMP}{Internet Control Message Protocol}
\acrodef{FTP}{File Transfer Protocol}
\acrodef{CE}{coverage enhancement}
\acrodef{PRB}{physical resource block}
\acrodef{LTE-A}{LTE-Advanced}
\acrodef{UL}{uplink}
\acrodef{DL}{downlink}
\acrodef{MIB}{master information block}
\acrodef{SIB}{system information block}
\acrodef{GSM}{Global System for Mobile Communications}
\acrodef{FDD}{Frequency Division Duplex}
\acrodef{OFDMA}{Orthogonal Frequency Division Multiple Access}
\acrodef{SC-FDMA}{Single Carrier Frequency Division Multiple Access}
\acrodef{UE}{user equipment}
\acrodef{eDRX}{extended discontinuous reception}
\acrodef{PSM}{power saving mode}
\acrodef{RAN}{radio access network}

\begin{document}	
\title{Coverage and Deployment Analysis of Narrowband Internet of Things in the Wild}

\author{Konstantinos~Kousias, Giuseppe~Caso, Özgü~Alay, Anna~Brunstrom, Luca~De~Nardis, Maria-Gabriella~Di~Benedetto and Marco Neri
\thanks{K. Kousias is with Simula Research Laboratory.}
\thanks{G. Caso is with Simula Metropolitan CDE.}
\thanks{Ö. Alay is with University of Oslo and Simula Metropolitan CDE.}
\thanks{A. Brunstrom is with Karlstad University.}
\thanks{L. De Nardis and M.-G. Di Benedetto are with Sapienza Uni. of Rome.}
\thanks{M. Neri is with Rohde\&Schwarz.}
}


\maketitle

\begin{abstract}

Narrowband Internet of Things (NB-IoT) is gaining momentum as a promising technology for massive Machine Type Communication (mMTC). 
Given that its deployment is rapidly progressing worldwide, measurement campaigns and performance analyses are needed to better understand the system and move toward its enhancement.
With this aim, this paper presents a large scale measurement campaign and empirical analysis of NB-IoT on operational networks, and discloses valuable insights in terms of deployment strategies and radio coverage performance. 
The reported results also serve as examples showing the potential usage of the collected dataset, which we make open-source along with a lightweight data visualization platform.

\end{abstract}

\IEEEpeerreviewmaketitle
\section{Introduction}
\label{Introduction}

As part of beyond-4G and 5G systems, the \ac{MTC} paradigm provides the ideal substrate toward cellular \ac{IoT}, and is leading to a shift in cellular network design and deployment. On the one hand, \ac{MTC} introduces a novel degree of heterogeneity, given that the \emph{things}, i.e., autonomous devices with novel features and requirements, must be integrated into the mobile network, which was originally designed for serving humans with their peculiar traffic. On the other hand, it requires long term 
performance analyses for stable 
network deployment, in particular when its \emph{massive} nature (mMTC), in terms of the unprecedented number of devices, is considered \cite{dhillon2017wide}\cite{liberg2019cellular}.

A relevant step toward enabling mMTC is represented by the 2016 Release 13 (Rel-13) standard by the 3rd generation partnership project (3GPP), where three technologies were proposed: Extended Coverage \ac{GSM} \ac{IoT} (EC-GSM-IoT), \ac{LTE} for \ac{MTC} (LTE-M), and \ac{NB-IoT} \cite{liberg2019cellular}. These represent the cellular options for so-called low power wide area networks (LPWANs), which aim to deliver massive \ac{IoT} services over wide areas, i.e., 
several kilometers, with low costs and power consumption \cite{raza2017low}.

Since Rel-13, and considering the advances in Rel-14 (2017) and Rel-15 (2018), \ac{NB-IoT} is triggering significant attention across researchers and operators as one of the most appealing LPWANs 
\cite{wang2017primer}\cite{yang2017narrowband}\cite{chen2017narrow}. 
Hence, the theoretical aspects of \ac{NB-IoT} are being 
formalized and analyzed, while several mobile operators are launching and making operational \ac{NB-IoT} networks worldwide \cite{GSMA_report_2019}.

As the \ac{NB-IoT} deployment is progressing at a rapid pace, field trials and measurement campaigns 
become of extreme interest, considering that these are the very first attempts of enabling \ac{IoT} services on the cellular architecture, and thus a closer empirical look is needed to better understand and optimize the system. 
To this end, data-driven analyses are crucial for both researchers and operators, as they allow to identify correlations and causalities between deployment choices and performance, highlight encountered challenges, and derive guidelines for further research and development. However, extensive measurement campaigns are often scarcely available to researchers, and rather costly and time consuming for operators, which may thus opt for less expensive but sub-optimal alternatives. These include simulation-based studies \cite{lauridsen2017coverage}, which provide general analyses that cannot perfectly match with real scenarios. 

Considering the above motivations, this paper presents a large scale measurement campaign of \ac{NB-IoT} coverage for two Norwegian and three Italian operators, conducted in the cities of Oslo and Rome during 2019. To the best of our knowledge, this represents the first empirical analysis of \ac{NB-IoT} performance on operational networks, which considers coverage aspects across heterogeneous scenarios and environments. The main contributions of this paper are:
\begin{itemize}
    \item We conduct an analysis of the strategies being adopted for deploying \ac{NB-IoT}, aiming to highlight implementation trends and derive takeaways for improvement. 
    \item We assess \ac{NB-IoT} coverage, depicting how the current system deployment reflects in service availability across urban scenarios. 
    \item We conduct a comparison between two \ac{NB-IoT} spectrum operation modes, \emph{in-band} and \emph{guard-band}, to understand \ac{NB-IoT}'s coexistence with \ac{LTE}. 
    \item We open-source our dataset, comprising of \ac{NB-IoT} and \ac{LTE} coverage measurements in Oslo and Rome, 
    to support the discovery of new insights and research. We also provide a web platform for geo-referenced visualization of the collected data \cite{nbiotweb}.  
\end{itemize}

The article is organized as follows: a brief description of \ac{NB-IoT} technology is first provided, followed by an overview of the experimental design. 
We then discuss our findings and conclude our work.  

\section{Technology Description}
\label{Background}
\ac{NB-IoT} is a radio interface implemented over the cellular licensed spectrum. It offers high deployment flexibility and integration with the existing architecture, minimizing costs and complexity at network and device sides, and providing performance in line with mMTC expectations.
Below, we describe \ac{NB-IoT} operation modes, possible deployment strategies, and coverage aspects, which are the focus of this paper. Moreover, we mention the main features in Rel-13 \cite{wang2017primer}, since this release is being mostly deployed, as confirmed by our measurement campaign. We refer the reader to \cite{liberg2019cellular}\cite{hoglund2017overview} 
for the analysis of \ac{NB-IoT} advances in Rel-14 and Rel-15.

\textit{\textbf{Operation modes and deployment strategies}}: 
\ac{NB-IoT} devices operate over either a 200 kHz \ac{GSM}-like channel or an \ac{LTE} \ac{PRB} of 180 kHz, allowing coexistence with both \ac{GSM} and \ac{LTE}. Three different operation modes are defined: 
\begin{itemize}
    \item \emph{stand-alone}, which uses a 200 kHz channel obtained by refarming the \ac{GSM} spectrum,  
    \item \emph{in‐band}, which uses a single \ac{PRB} within a set of \acp{PRB} used by \ac{LTE}, selected in order to minimize interference from/to \ac{LTE}, and
    \item \emph{guard-band}, which leverages a \ac{PRB} within a guard band among different sets of \ac{LTE} \acp{PRB}.
\end{itemize}
After selecting one of the modes, the operators can provide \ac{NB-IoT} services via a software upgrade of their infrastructure, i.e., enhancing the capabilities of their \acp{eNB} and cells, at least in areas where these are already present for serving broadband users. 

The operators can select several deployment strategies and differently leverage the trade-off between costs and performance. In particular, given a broadband area, e.g., an urban environment, operators may either activate an \ac{NB-IoT} carrier in all already deployed \ac{LTE} \acp{eNB} and cells, or select some of them. The first option is somehow simpler, since it does not require in-depth analysis for coverage optimization. However, it may increase the operational costs, including energy consumption, and result in redundant deployment when \ac{NB-IoT} use cases and \ac{CE} techniques are considered (see later in this section). The second option requires a more careful planning, but may lead to a better trade-off between costs and \ac{QoS}. 

Specifically for the in-band mode, the coexistence with \ac{LTE} is another aspect to consider when deciding between the two aforementioned options. On the one hand, \ac{NB-IoT} activation in specific \acp{eNB}/cells leads to possible interference from/to \ac{LTE}, since the \ac{LTE}-only \acp{eNB}/cells may use the \ac{PRB} dedicated to \ac{NB-IoT} for their traffic. On the other hand, \ac{NB-IoT} full deployment leads to sub-optimal resource usage, since a specific \ac{PRB} is exclusively dedicated to infrequent and sporadic \ac{NB-IoT} traffic, at the expenses of \ac{LTE} end-users. Moreover, in areas with low-to-null broadband coverage, e.g., rural and deep indoor environments, the operators may a-priori install dedicated but costly \acp{eNB}, or first check whether \ac{NB-IoT} \ac{CE} techniques allow the reuse of the existing infrastructure, triggering the installation only in cases of negative response.

\textit{\textbf{\ac{CE} techniques}}: 
\ac{NB-IoT} targets service reliability and delay-tolerant \ac{UL} data exchange. Hence, advanced modulation and coding schemes are not supported. 
Rather, \ac{CE} techniques are used, aiming to favour connectivity in harsh environments, such as dense urban and deep indoor. A first \ac{CE} effect is obtained by narrowing down the bandwidth with respect to \ac{LTE}, since this focuses the transmitted power on smaller bands, at the cost of reducing the data rate. Moreover, \ac{NB-IoT} standards allow repeated transmissions, which increase the probability of correct reception. In particular, \ac{DL} and \ac{UL} signals can be repeated up to 2048 and 128 times, respectively. The number of repetitions depends on radio conditions and operator configurations, these latter transmitted in \ac{MIB} and \ac{SIB} messages. Repetition settings are given in \ac{SIB}2 messages. 

The device estimates its coverage conditions during the \ac{RA} procedure, which triggers the connection to a cell, i.e., the one detected with highest \ac{RSRP} [dBm]. The comparison between \ac{RSRP} and operator-defined thresholds allows to estimate a \ac{CL} \cite{3GPP_ts36133}. Up to two thresholds can be defined, leading to three possible \acp{CL}; \ac{CL}0 represents \ac{LTE}-like radio conditions, while \ac{CL}1 and \ac{CL}2 apply to challenging scenarios requiring more repetitions. During consecutive \ac{RA} attempts the device can adjust its \ac{CL} estimate and move to higher \acp{CL}, if it experiences connection failures in the first attempts.

\textit{\textbf{Other Features}}: 
\ac{DL} and \ac{UL} resources are accessed in \ac{FDD} mode. \ac{OFDMA} is applied in \ac{DL}, with 15 kHz subcarrier spacing. 
The \ac{PRB} is divided into seven OFDM symbols of twelve subcarriers each, and occupies 0.5 ms. 
\ac{SC-FDMA} is applied in \ac{UL}, with a subcarrier spacing of 15 kHz or 3.75 kHz. 

\ac{NB-IoT} devices can be \emph{idle} and \emph{connected}. Idle devices are not \emph{functionally} connected to the network, and actuate the procedures for switching into connected, including cell selection and tracking of control messages (\emph{paging monitoring}). 
After selecting a cell, the devices transit from idle to connected via \ac{RA},  
and exchange data while continuing paging monitoring. 

Targeting energy efficiency and long device battery lifetime, \ac{NB-IoT} standards introduce 
(i) \emph{\ac{eDRX}}, which allows to perform paging monitoring more infrequently compared to \ac{LTE}, 
and (ii) \emph{\ac{PSM}}, which allows idle devices to disconnect the radio and minimize energy consumption \cite{martinez2019exploring}. 

\textit{\textbf{Comparison with other LPWANs:}}
\ac{NB-IoT} plays a leading role across LPWANs, with constantly increasing market shares \cite{raza2017low}\cite{popli2018survey}. Compared to other 3GPP technologies, \ac{NB-IoT} shares some features and use cases with EC-GSM-IoT, but provides lower device complexity and better integration with \ac{GSM}, \ac{LTE}, and 5G. \ac{NB-IoT} and LTE-M target complementary applications, with LTE-M having higher device complexity and costs. Considering LPWANs in the unlicensed spectrum, \ac{NB-IoT} has a competitor in Long Range Wide Area Network (LoRaWAN). 
They provide comparable throughput and latency performance, and adopt different security mechanisms. \ac{NB-IoT} outperforms LoRaWAN in terms of communication reliability, due to the use of licensed spectrum and well-established infrastructures \cite{liberg2019cellular}. 

\section{Experimental Design}
\label{Experimental_Design}

In the above overview, we highlight \ac{NB-IoT} deployment and coverage aspects discussed in this paper.
We now present our measurement campaigns and analyses. 
Particularly, in this section we describe the adopted hardware and software components, and summarize experimental setup and collected dataset. 
Finally, we introduce our data visualization framework.

\subsection{Measurement System}
\label{Measurement_System}

For the \ac{NB-IoT} measurements in Oslo and Rome, we used the Rohde\&Schwarz (R\&S) TSMA6 toolkit, an Exelonix Narrowband (NB) USB device, and a \ac{GPS} antenna. 
TSMA6 is a system integrating: 
\begin{itemize}
\item A spectrum scanner, for passive measurement of 3GPP technologies up to 6 GHz, including 5G New Radio (NR). It supports \ac{NB-IoT} signal detection and decoding in in-band, guard-band, and stand-alone.   
\item A laptop, where the controlling software, named ROMES4, is installed. Combined with scanner and device, i.e., the Exelonix module in our case, and exploiting \ac{GPS} georeference, ROMES4 provides an overview of coverage, interference, and \ac{QoS} measurements.
\end{itemize}

We leveraged two further features from TSMA6, i.e., the \emph{automatic channel detection}, which performs automatic detection of active channels for all technologies in the specified spectrum, and the \emph{base transceiver station position estimation}, which combines passive measurements and \ac{GPS} to estimate the position of the cells forming the operators' infrastructures.

The Exelonix module is a Qualcomm-based device supporting \ac{NB-IoT} and \ac{LTE}-M. 
We embedded the module with \ac{NB-IoT} SIM cards from the operators under test, and used it to monitor the radio conditions of the serving cell, and execute repeated connections to the operators' networks.
Hence, we are able to analyze the \ac{RA} procedure, including the \ac{CL} estimation, aiming to reveal further aspects related to operator-specific configurations, e.g., how the \ac{RSRP} thresholds adopted in the \ac{CL} estimation impact the perceived coverage.

\subsection{Measurement Campaigns}
\label{Measurement_Campaigns}


We performed two measurement campaigns. 
The first campaign was designed to explore city-wide coverage and deployment aspects under heterogeneous scenarios, and covered a period of three weeks during summer 2019 in Oslo, Norway.
We enabled the scanner to perform passive measurements on four \ac{LTE} bands (including guard bands), i.e., Band 1, 3, 7, and 20, and detected three \ac{LTE} operators, denoted in the following as 
Op\textsubscript{k,N}, where 
k identifies the operator and N stands for Norway. 
To guarantee reliability and completeness, we conducted measurements in various city areas and different scenarios, that we label as \emph{\ac{DI}} (14), for basements and deep enclosed spaces, \emph{\ac{I}} (48), for houses and multi-floor buildings, \emph{\ac{OW}} (8), for outdoor while walking, and \emph{\ac{OD}} (14), for outdoor while on public transport.
Numbers in parenthesis represent the number of sub-campaigns for each scenario.
We replicated a subset of our measurements over time (i.e., morning vs. afternoon vs. evening, and week vs. weekend), to account for temporal effects.
In the second campaign we collected a smaller dataset composed of 3 sub-campaigns (one for \emph{I} and two for \emph{OD} scenarios), within a couple of days of 2019 in Rome, Italy, to compare the performance between in-band and guard-band modes. 
The dataset features measurements related to three operators (Op\textsubscript{k,I}, where
k identifies the operator and I stands for Italy) in Band 20.
At the time of the collection, Op\textsubscript{2,I} and Op\textsubscript{3,I} were deploying \ac{NB-IoT} in guard-band, while Op\textsubscript{1,I} was testing the in-band option. 
This makes the dataset fitting for a comparison between the two modes. 
We report that, as confirmed by following tests, Op\textsubscript{1,I} has moved toward a guard-band deployment. 
However, the dataset remains valid for comparing the two modes. 


The complete dataset consists of 1.2M \ac{LTE} and 1.4M \ac{NB-IoT} passive scans for Oslo, and 121K \ac{LTE} and 51K \ac{NB-IoT} passive scans for Rome.
The list of collected attributes is provided in \cite{nbiotweb}. 
To anonymize the operators' identity, \ac{MNC}, \ac{EARFCN}, and \ac{CID} are given as references and not associated to real values.

\subsection{Visualization}
\label{Visualization}
Designing and implementing a platform that enables interactive geo-spatial visualization is beneficial for 
discovering operators' \acp{eNB} spatial deployment and pinpointing at a glance areas with limited coverage. 
Thereby, we design an open-source visualization platform showing \ac{eNB} placement and coverage for each operator under test \cite{nbiotweb}. 
We implement the platform using an \textit{R} interface to \textit{Leaflet}, 
an open-source JavaScript library for mobile-friendly 
maps. 
Users can exploit several features, from controlling which layers they see on the map to dynamically altering the observed coverage based on the zoom level. 
We include additional plugins and add-ons to enhance end-user experience. 

\section{Performance Evaluation}
\label{Performance}

The deployment of the cellular \ac{RAN} is driven by the need of optimizing the coverage and making the service accessible to end-users. 
This aspect is more challenging for cellular \ac{IoT} technologies, as they are expected to mostly leverage the existing infrastructure, which is however tailored for broadband services. 
In this section, we present the results of our  measurement analysis, which contrast deployment strategies and coverage performance for the two operators currently providing \ac{NB-IoT} in Oslo.
We also study the implications of deploying \ac{NB-IoT} in in-band or guard-band, by leveraging the dataset collected in Rome. 

\subsection{Network Deployment Strategy}
\label{Network_Deployment_Strategy}

\begin{figure*}[t]
	\includegraphics[keepaspectratio,width = 0.7\linewidth]{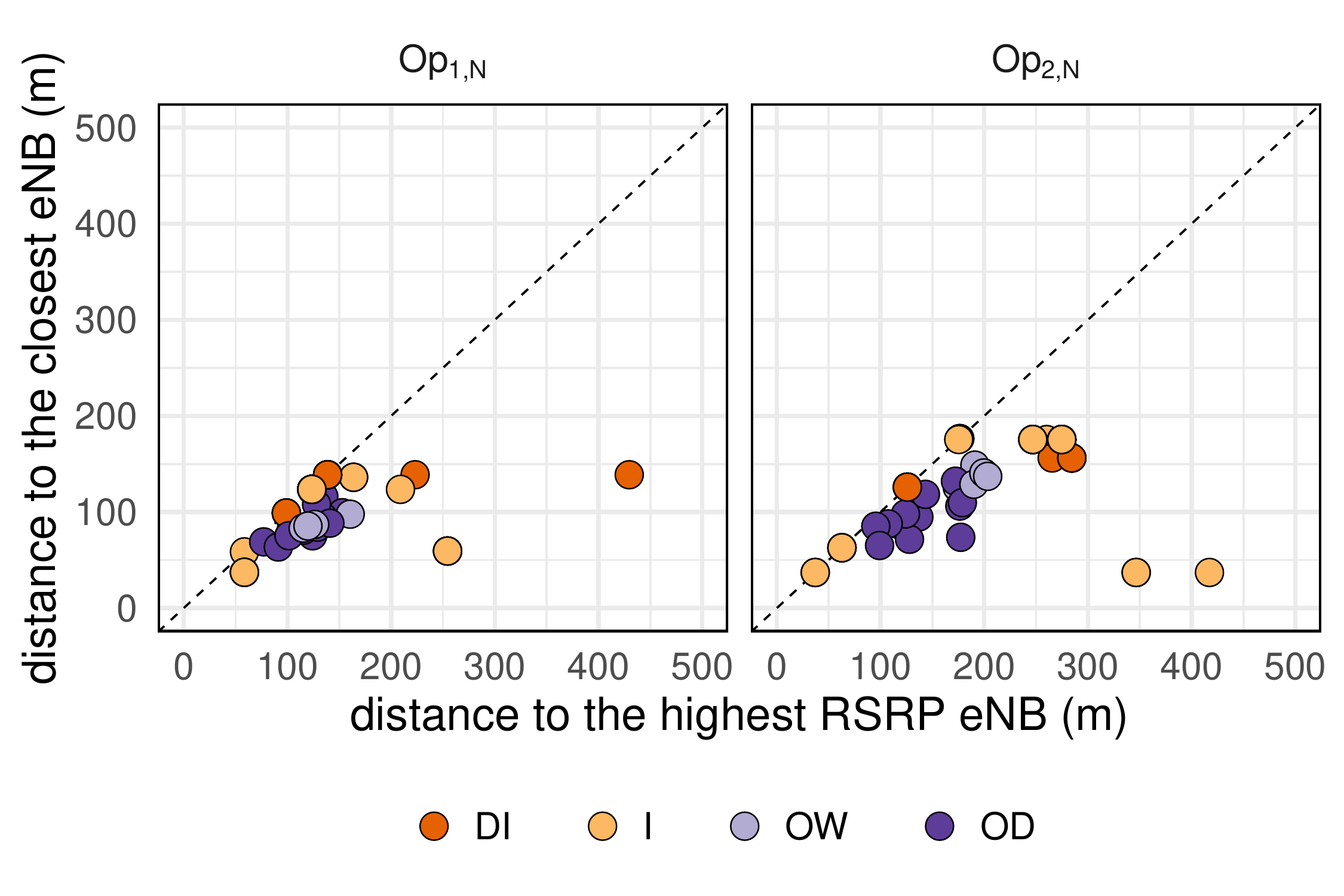}
	\centering
    \caption{\small Scatterplot between average distance to the \ac{eNB} with highest \ac{RSRP} 
    and average distance to the nearest \ac{eNB} 
    for Op\textsubscript{1,N} (left) and Op\textsubscript{2,N} (right). Different colors represent different scenarios.} 
	\label{fig:scatter}
	\vspace{-4mm}
\end{figure*}

\ac{RAN} deployment is a challenging optimization task \cite{wang2015efficient}.
The targets of the operators include (i) to ensure sufficient coverage, (ii) to satisfy \ac{QoS} requirements, and (iii) to efficiently deal with energy and cost constraints.
Hence, they aim to optimize the \ac{eNB} placement by considering environmental characteristics, i.e., density and structure of surrounding buildings.
However, business expenses,
radiation safety levels, and interference between cells, are some aspects that limit the qualified spots for installing an \ac{eNB}, hence favoring alternative locations. Next, we empirically analyze the deployment strategies for operators providing \ac{NB-IoT} in Oslo.

\textit{\textbf{Deployment statistics:}} Table \ref{table:stats} provides per-operator statistical insights with respect to the number of detected \acp{eNB} and \acp{EARFCN} used for \ac{NB-IoT} and \ac{LTE}.  
We observe that, across the monitored bands, Op\textsubscript{1,N} and Op\textsubscript{2,N} have activated one \ac{NB-IoT} carrier each in the guard bands of Band 20.
Considering the infrastructure, Op\textsubscript{1,N} is dominant in terms of number of \acp{eNB} for both technologies, 
implying a denser deployment compared to Op\textsubscript{2,N}. 
Both operators leverage the existing \ac{LTE} infrastructure for deploying \ac{NB-IoT}, with no additional \acp{eNB} installed. 
In this regard, nearly 86\% of the detected \ac{LTE} \acp{eNB} have been reconfigured for \ac{NB-IoT}.  
The few \ac{NB-IoT}-only \acp{eNB} detected for Op\textsubscript{1,N} could be explained by considering the more penetrating nature of \ac{NB-IoT}, or should be appointed to other causes that prevented \ac{LTE} detection.
We also observe that the \acp{eNB} not supporting \ac{NB-IoT} are in 90\% of the cases operating at a band other than Band 20. 
This indicates that almost all \acp{eNB} operating in Band 20 support \ac{NB-IoT}.
Finally, we highlight that the operators leverage a different number of \acp{EARFCN} for \ac{LTE}, while only one for \ac{NB-IoT}.

\begin{table}
\centering
\caption{\small Network deployment statistics with regard to number of \acp{eNB} and \acp{EARFCN} per technology.
\acp{eNB}\% is defined as the ratio between the number of \ac{LTE}/\ac{NB-IoT} \acp{eNB} and the total number of \acp{eNB}.
Absolute numbers are provided in parenthesis. 
}
\begin{tabular}{|l|c|c||c|c|}
\hline
& \multicolumn{2}{c||}{\acp{eNB}\%} & \multicolumn{2}{c|}{\acp{EARFCN}} \\
 \hline
 &LTE&NB-IoT&LTE&NB-IoT\\
 \hline
  Op\textsubscript{1,N}&96.6\% (167)&84.3\% (146)&6&1\\
 Op\textsubscript{2,N}&100\% (122)&87.7\% (107)&4&1\\
 Op\textsubscript{3,N}&100\% (70)& NA &2& NA \\
 \hline
\end{tabular}
\label{table:stats}
\vspace{-4mm}
\end{table}


\textit{\textbf{Deployment optimality:}} 
\ac{RSRP} is a key metric for handover and cell (re-)selection phases, hence, it is a critical parameter when evaluating how radio coverage is affected by the network infrastructure.
In the following, we evaluate \emph{deployment optimality}, which assumes given a location, the closest \ac{eNB} would offer the highest \ac{RSRP} under ideal propagation and environmental scenarios, and also assuming constraintless \ac{eNB} placement. 
In practice though, several factors may inhibit this situation, particularly in dense urban environments, such as multipath propagation, network congestion and interference, and constrained \ac{eNB} placement.  

Figure \ref{fig:scatter} shows whether the operators are close to an optimal \ac{NB-IoT} deployment in Oslo. 
For each location in a sub-campaign, we compute (i) the distance toward the \ac{eNB} detected with highest \ac{RSRP}, 
and (ii) the distance toward the nearest \ac{eNB}, 
both in meters. 
Then, for each sub-campaign, we average  across all locations. 
In an optimal deployment situation, we expect a linear relationship between the two distances.
Thereby, the deviation from the diagonal represents how far the deployment refrains from being optimal.   
We observe that both operators approach deployment optimality in several indoor scenarios, while slightly deviate in outdoor sub-campaigns. In particular, Op\textsubscript{1,N} mostly works in a short distance regime, i.e., less than 150 meters, due to its dense infrastructure. Op\textsubscript{2,N} deviates from the identity line more frequently, and several sub-campaigns present distances exceeding 150 meters, thus hinting sub-optimal deployment. For both operators, a negative joint impact of propagation conditions and deployment sub-optimality is highlighted by large deviations observed for specific \ac{DI} and \ac{I} sub-campaigns.

\begin{figure*}[t]
	\includegraphics[keepaspectratio,width = 0.7\linewidth]{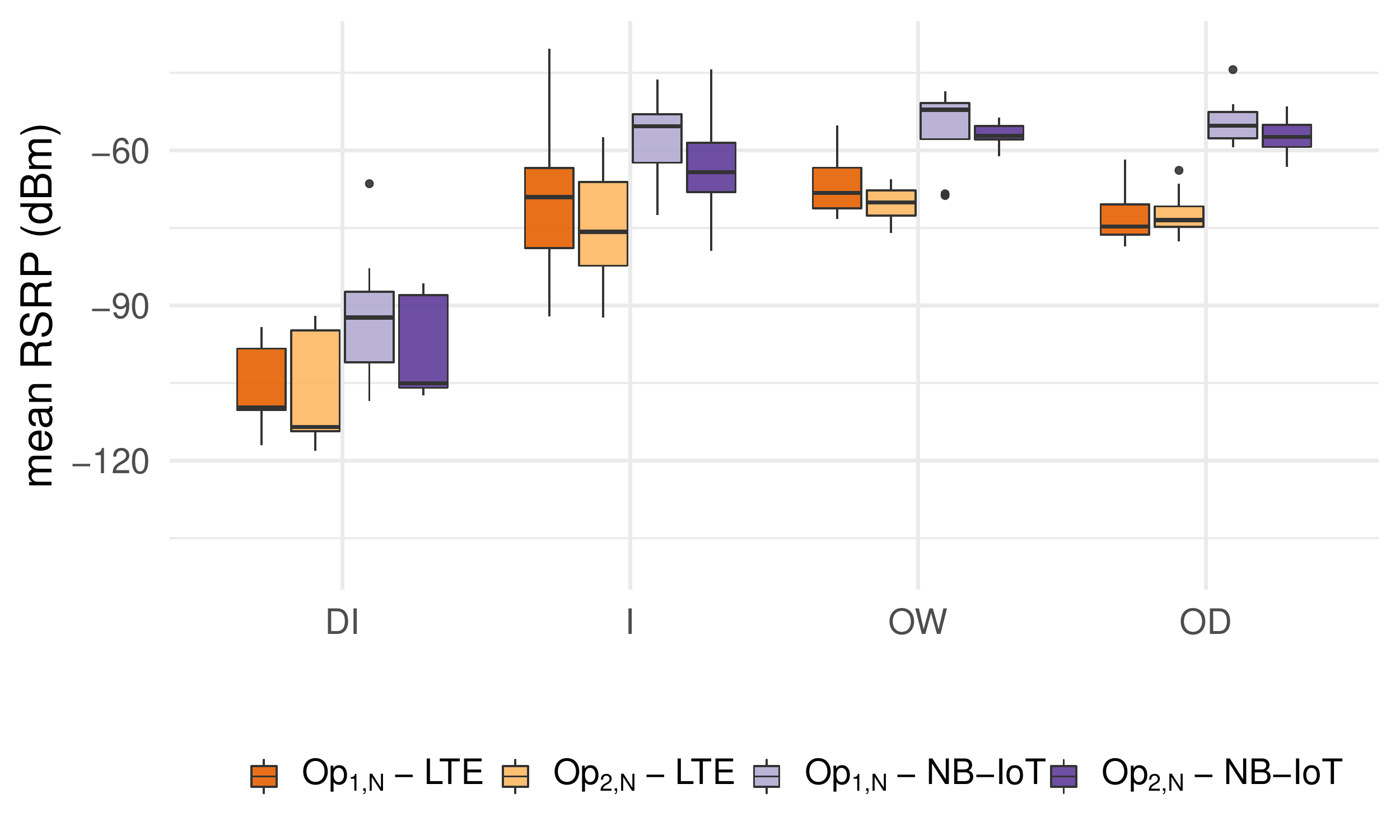}
	\centering
    \caption{\small Sub-campaign average coverage in terms of \ac{RSRP} [dBm],
    grouped by scenario and divided by a combination of operator (Op\textsubscript{1,N}, Op\textsubscript{2,N}) and technology (\ac{LTE}, \ac{NB-IoT}).
    }
	\label{fig:rsrp}
\end{figure*}

\textit{\textbf{Takeaways:}}
In initial deployment phases, the operators are actuating rather \emph{dense} \ac{NB-IoT} deployment strategies, with large amounts of pre-existing \acp{eNB} now supporting \ac{NB-IoT}. This solution is sub-optimal in terms of operational costs, and leads to increased carbon emissions, for which the \ac{RAN} is already the main contributor across network functions \cite{popli2018survey}. The analysis of real deployments can support the derivation of optimization strategies, e.g., dynamic (de-)activation of specific \acp{eNB},  
moving toward \emph{green} cellular \ac{IoT}.

\subsection{Radio Coverage}
\label{Radio_Coverage}

We now analyze \ac{NB-IoT} coverage for both operators, showing how it changes across scenarios, and exploiting the \ac{LTE} dataset for comparison. 
As above, due to its importance in cellular systems, we consider \ac{RSRP} as a key indicator. 
In particular, for each measurement location in a sub-campaign, the coverage for an operator is defined as the highest \ac{RSRP} perceived among all the \acp{CID} detected for that operator. 
We then express the sub-campaign average coverage, 
by averaging the \ac{RSRP} across locations.  
 

\textit{\textbf{Technology and scenario comparison:}} Figure \ref{fig:rsrp} depicts the distribution of average \ac{RSRP} in a boxplot format with sub-campaigns grouped per scenario and colored by operator and technology. We validated the statistical significance by leveraging the Kruskal-Wallis and Dunn's tests, aiming to identify which distributions have statistically different mean values. 
Due to space limitations, we report the results in \cite{nbiotweb}.

We observe that, compared to \ac{LTE}, \ac{NB-IoT} provides statistically significant \ac{RSRP} boosts of 11.73, 12.29, 12.06 and 16.71 dB on average for each scenario, respectively. 
This result is in line with the power boosting expected by 3GPP TS 36.104 \cite{3GPP_ts36104}, which is at least +6 dB when evaluated as the difference between the power of the entire \ac{NB-IoT} carrier (180 kHz) and the average power over all carriers (\ac{LTE} and \ac{NB-IoT}). 


We also compare \ac{NB-IoT} average \ac{RSRP} across scenarios.
In particular, a statistically significant 
increase of 36.36 and 35.70 dB for Op\textsubscript{1,N} and Op\textsubscript{2,N}, respectively, is observed when comparing \ac{I} with \ac{DI} scenarios. 
This shows the negative effect of \ac{DI} on signal propagation, which needs to be compensated by \ac{CE} techniques. 
The deviation between outdoor scenarios to \ac{I} is instead reduced, with an 
average increase of 1.43 dB for Op\textsubscript{1,N} and  5.82 dB 
for Op\textsubscript{2,N}. 
Comparing the operators,  Op\textsubscript{1,N} consistently provides better \ac{NB-IoT} coverage (3.95 dB on average, and statistically significant for the \ac{I} scenario), which ties back to the results on deployment statistics and optimality.

\renewcommand{\tabcolsep}{1.5pt}
\begin{table}
\centering
 \caption{\small Ratio of being in a specific \ac{CL}, grouped by scenario, for Op\textsubscript{1,N} and Op\textsubscript{2,N}. The ratio is evaluated as the number of readings for a \ac{CL} divided by the number of readings for all \acp{CL}.}
\begin{tabular}{||c||c|c|c||c|c|c||c|c|c||c|c|c||}
\hline
Operators & \multicolumn{12}{c||}{Scenarios \& \acp{CL}} \\
\hline\hline
& \multicolumn{3}{c||}{DI} & \multicolumn{3}{c||}{I} & \multicolumn{3}{c||}{OW} & \multicolumn{3}{c||}{OD} \\
 \hline
&\ac{CL}0&\ac{CL}1&\ac{CL}2& \ac{CL}0 & \ac{CL}1 & \ac{CL}2 & \ac{CL}0 & \ac{CL}1 & \ac{CL}2 & \ac{CL}0 & \ac{CL}1 & \ac{CL}2\\
 \hline
 Op\textsubscript{1,N} &.72&.13&.15 & .99 & .005 & .005 & .86 & .09 & .05 & .84 & .08 & .08\\
 Op\textsubscript{2,N} &.16&.77&.07& .89 & .105& .005 & .78 & .17 & .05 & .80 & .15 & .05\\
 \hline
\end{tabular}
\label{table:CL}
\end{table}

\begin{figure*}[t]
	\includegraphics[keepaspectratio,width = 0.7\linewidth]{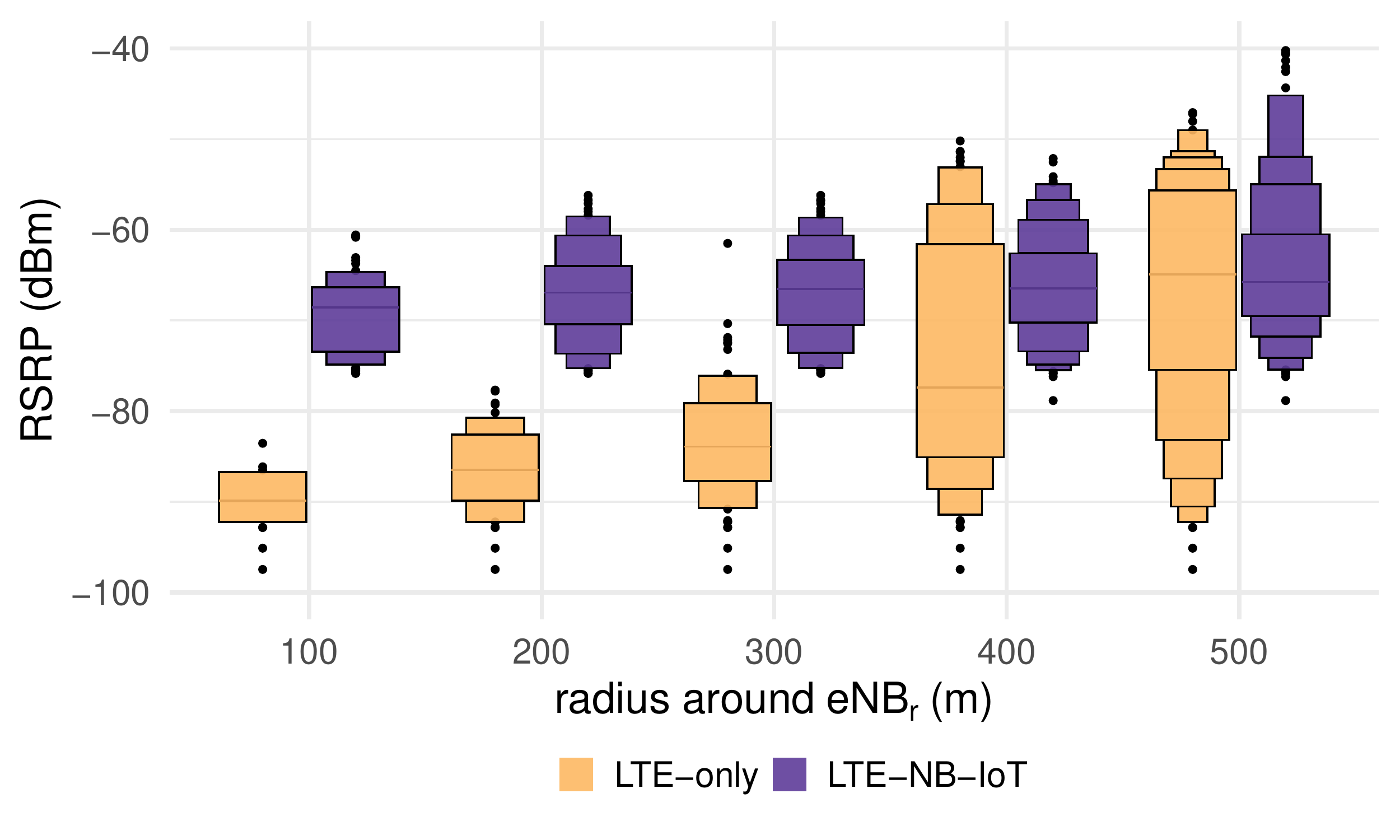}
	\centering
    \caption{\small Network deployment performance evaluation in terms of \ac{RSRP} [dBm] for the in-band Op\textsubscript{1,I} deployments.} 
	\label{fig:letter}
\end{figure*}

\textit{\textbf{Coverage Levels:}} To better understand how coverage is affected by operators' configurations, we report in Table \ref{table:CL} the ratio of being in a specific \ac{CL}, grouped per scenario and split by operator.
We retrieve the \ac{CL} readings by monitoring via TSMA6 the \ac{RA} attempts performed by the Exelonix module. We then evaluate the ratio as the number of readings for a \ac{CL} divided by the total number of \ac{CL} readings. 
By decoding SIB2 messages, we observe that the operators apply different \ac{RSRP} thresholds for the \ac{CL} estimation, 5 dB more conservative for Op\textsubscript{2,N}, which is thus more likely to work at higher \acp{CL}. 
When combined with \ac{RSRP} values of the serving \ac{CID}, the different thresholds explain the main results in Table \ref{table:CL}. 
First, Op\textsubscript{2,N} has in general lower coverage than Op\textsubscript{1,N}, and thus tries to enhance \ac{NB-IoT} reliability by operating at higher \acp{CL} more, at the cost of higher energy consumption and more repetitions. 
This is evident in \ac{DI}, where Op\textsubscript{2,N} works predominantly in \ac{CL}1 while Op\textsubscript{1,N} exhibits a high ratio of operating at \ac{CL}0. 
Second, operations at \ac{CL}0 are predominant in \ac{I}, \ac{OW}, and \ac{OD} scenarios, showing that, in generic urban environments, \ac{NB-IoT} is likely to work at radio conditions similar to \ac{LTE}. 
Finally, we observe that outdoor scenarios increase the ratio of being at higher \acp{CL} when compared to the indoor case, even though higher average \ac{RSRP} is observed for the former (Figure \ref{fig:rsrp}).
We argue that this effect is due to the higher heterogeneity of the outdoor scenarios, in terms of radio and mobility aspects. 
The result suggests that in more heterogeneous and dynamic scenarios, the current \ac{CL} estimate procedure may be sub-optimal and cause several attempts before the \ac{RA} succeeds, leading to energy inefficiency. 

\textit{\textbf{Takeaways:}}
\ac{NB-IoT} results in significant coverage improvements with respect to \ac{LTE}, but operators' configurations have a direct impact on how devices perceive the coverage, execute their operations, and perform in terms of \ac{QoS}. Empirical data can be used to refine the models adopted in previous work \cite{lauridsen2017coverage}, 
resulting in better modeling \ac{NB-IoT} coverage and propagation which can pave the way to optimize deployments, configurations, and \ac{QoS}.

\subsection{In-band and Guard-band deployment}
\label{Guard-band_and_In-band_deployment}
In this subsection, we perform a comparison between in-band and guard-band modes, 
quantifying their impact on coverage. 
As mentioned before, the in-band mode may challenge \ac{NB-IoT}/\ac{LTE} coexistence in case of partial \ac{NB-IoT} deployment across the \ac{LTE} infrastructure. 



We hence focus on in-band mode, and look into the Italian dataset to find a situation of partial \ac{NB-IoT} deployment for operator Op\textsubscript{1,I}. 
Our goal is to study the characteristics of the \ac{NB-IoT} signal around an \ac{LTE}-only \ac{eNB}, denoted as eNB\textsubscript{r}, toward discovering and quantifying potential interference from the \ac{LTE} signal.
Hence, we consider eNB\textsubscript{r} and a \ac{NB-IoT}-enabled \ac{eNB}, denoted as eNB\textsubscript{x}. 
We then draw two circles  
around eNB\textsubscript{r} and eNB\textsubscript{x}, and isolate all \ac{RSRP} readings from eNB\textsubscript{x}, captured in the intersection of the two.
We repeat the process by increasing the radius around eNB\textsubscript{r},  
and appending new \ac{RSRP} readings each time. 
We compare this scenario with a similar topology for Op\textsubscript{1,I}, where, however, both eNB\textsubscript{r} and eNB\textsubscript{x} are \ac{NB-IoT}-enabled. 
To provide a fair comparison, 
we select almost symmetric configurations, with analogous distance between the two \acp{eNB}. 
Last, we repeat the same analysis for the guard-band operator Op\textsubscript{2,I}.

Figure \ref{fig:letter} shows the \ac{NB-IoT} \ac{RSRP} distribution as a function of the radius around eNB\textsubscript{r}  
between 100 and 500 meters 
(the radius around eNB\textsubscript{x} is fixed to the distance between the two \acp{eNB}), grouped by scenario, i.e., \ac{LTE}-only vs. \ac{LTE}-\ac{NB-IoT} eNB\textsubscript{r}, for an in-band 
deployment of Op\textsubscript{1,I}.
We observe that 
the effect of interference from \ac{LTE} is visible especially in close proximity to eNB\textsubscript{r}, while it diminishes as the radius  
increases and finally vanishes at around 500 meters.
Contrarily, we observe no interference impact at different radii for the guard-band deployment of Op\textsubscript{2,I},
confirming that there is no visible interference for the guard-band deployment. 
We observe similar trends for the guard-band deployment of Op\textsubscript{3,I}.

In the supplementary material \cite{nbiotweb}, we show the analyzed topology, present guard-band results and also validated the statistical significance of the results.


\textit{\textbf{Takeaways:}}
The in-band mode poses coexistence challenges that need to be carefully considered. The empirical assessment of interference is key for the derivation of improved mitigation schemes, also in light of near future transition to 5G, which leads to further coexistence challenges \cite{popli2018survey}.

\section{Conclusion}
\label{Conclusions}

In this paper, we present the first publicly available measurement campaign and analysis of \ac{NB-IoT} on operational networks, focusing on aspects related to network deployment strategies and coverage. 
By leveraging the collected dataset, we first focus on network deployment, highlighting that empirical studies can pinpoint sub-optimal deployments and pave the way for better deployments with decreased costs and energy consumption. 
We then show that a dense reuse of the \ac{LTE} \ac{RAN} for deploying \ac{NB-IoT} results in a significant coverage increase with respect to \ac{LTE}, across heterogeneous scenarios and environments. 
We also analyze how operator-specific configurations affect end-devices' operations, leading to different estimates of the coverage quality. 
Finally, we assess the impact of in-band and guard-band modes in terms of \ac{LTE} interference, showing a non-negligible negative effect of the former, and pinpointing how empirical measurements can help to assess interference and derive mitigation schemes. 
The open-source nature of our dataset and visualization platform enables further data exploration toward the discovery of new insights and research perspectives.
\section{Acknowledgment}
\label{Acknowledgments}

This work is partially funded by the EU H2020 5GENESIS (815178), and by the Norwegian Research Council project MEMBRANE (250679).


\section*{Biographies}
\begin{IEEEbiographynophoto}{\textbf{Konstantinos Kousias}} 
    \small [S] is a PhD Student at the Faculty of Mathematics and Natural Sciences in University of Oslo affiliated with Simula Research Laboratory. 
    His research focuses on the empirical modeling and evaluation of mobile networks using data analytics and artificial intelligence. 
\end{IEEEbiographynophoto}

\begin{IEEEbiographynophoto}{\textbf{Giuseppe Caso}} 
    \small [M] is a Postdoctoral Fellow with the Mobile Systems and Analytics Department at Simula Metropolitan. He received the PhD degree from Sapienza University of Rome in 2016. 
    His research interests include cognitive communications, distributed learning, and IoT technologies.
\end{IEEEbiographynophoto}

\begin{IEEEbiographynophoto}{\textbf{Özgü Alay}} 
    \small [M] is an Associate Professor with the University of Oslo, Norway. She is also a Chief Research Scientist at Simula Metropolitan. Her research team focuses on the empirical characterization of mobile networks, and the design of protocols and applications for future mobile networks.
\end{IEEEbiographynophoto}

\begin{IEEEbiographynophoto}{\textbf{Anna Brunstrom}} 
    \small [M] is a Full Professor and Research Manager for the Distributed Systems and Communications Research Group at Karlstad University, Sweden. Her research interests include Internet architectures and protocols, techniques for low latency Internet communication, multi-path communication and performance evaluation of mobile broadband systems including 5G.  
\end{IEEEbiographynophoto}

\begin{IEEEbiographynophoto}{\textbf{Luca De Nardis}} 
    \small [M] (Ph.D. 2005) is an Assistant Professor in the Department of Information Engineering, Electronics and Telecommunications at Sapienza University of Rome.  
    His research interests focus on indoor positioning, UWB communications, cognitive radio, and routing protocols. 
\end{IEEEbiographynophoto}

\begin{IEEEbiographynophoto}{\textbf{Maria-Gabriella Di Benedetto}}
    \small [F] (Ph.D. 1987) is a Full Professor at the Department of Information Engineering, Electronics and Telecommunications at Sapienza University of Rome. 
    Her research interests include wireless communication systems, in particular impulse radio communications, and speech.
\end{IEEEbiographynophoto}

\begin{IEEEbiographynophoto}{\textbf{Marco Neri}}
	\small joined Rohde\&Schwarz in 2017 as Application Engineer for Mobile Network Testing. His focus is on $5$G and cellular IoT testing activities worldwide. 
\end{IEEEbiographynophoto}

\end{document}